\documentclass[manuscript]{aastex}
\usepackage{tikz}
\usepackage{amsmath}

\usepackage{graphicx}

\usepackage{pdflscape}

\slugcomment{\textbf{2025 June 10} }

\shorttitle{C/2021 O3}
\shortauthors{Jewitt, Li, Jaeger, Kim}

\begin{document}

\title{Down But Not Out: The Case of Long-Period Comet C/2021 O3 (Panstarrs)}

\author{
David Jewitt$^{1}$, Jing Li$^{1}$, Michael Jaeger$^{2}$ and Yoonyoung Kim$^{1}$
} 
\affil{$^1$Department of Earth, Planetary and Space Sciences, UCLA, 595 Charles Young Drive, Los Angeles, CA 90095\\
$^2$ Auf der Burg 225, A-3610 Weißenkirchen, Austria}

\email{djewitt@gmail.com}

\begin{abstract}
We combine ground- and space-based observations of long-period comet C/2021 O3 (Panstarrs) (perihelion distance 0.287 au) in order to investigate its reported near-perihelion destruction.  Pre-perihelion photometric observations show a remarkably small heliocentric dependence of the scattered light, $\propto r_H^{-s}$ with $s = 2.59\pm0.21$, distinct from values reported in other long-period comets, for which $s$ = 4 is the canonical standard. The  index is smaller than expected of coma production by equilibrium sublimation of either supervolatiles (for which $s \sim$ 4 is expected), or water ice ($s \sim$ 6 to 8) across the $\sim$4 au to 2 au range.  The absolute magnitude deduced from the pre-perihelion data is $H$ = 13.0$\pm$0.3 (coma scattering cross-section $\sim$225 km$^2$ for an assumed geometric albedo 0.04) while, after perihelion, the cross-section fades by a factor of 25 to $H$ = 16.5 ($\sim$9 km$^2$).  STEREO spacecraft observations near perihelion show a long debris trail whose properties are consistent with forward scattering from radius $\sim$7 $\mu$m particles.  The data show that the nucleus of C/2021 O3 was not destroyed at perihelion.  Although the lightcurve  from 3.9 au inbound to 0.8 au outbound cannot be uniquely interpreted, a simple and plausible explanation is provided by seasonal dimming on a nucleus having high obliquity and an asymmetric distribution of near-surface volatiles.  The survival of the nucleus against rotational disruption suggests a pre-perihelion nucleus radius $r_n \gtrsim$ 1.0 km while the photometric limit to the radius of the nucleus after perihelion is $r_n < 1.7$ km (geometric albedo 0.04 assumed). 
\end{abstract}


\section{INTRODUCTION}
\label{intro}
Comets are ice-containing bodies that likely formed between the orbits of the giant planets and which have been stored for the last 4.5 Gyr in one of two principal reservoirs. The Kuiper belt is the source of most short-period comets (periods $\le$200 years, most with sub-radian orbital inclinations and modest eccentricities) while the Oort cloud is the source of long-period comets (hereafter LPCs), with periods $>$200 years,  eccentricities near unity and an isotropic distribution of inclinations.  The LPCs themselves are notionally divided into two groups; the ``dynamically new LPCs'' have semimajor axes $|a| > 10^4$ au while LPCs with smaller $a$ are likely to have made previous visits to the planetary region and are sometimes called ``returning LPCs'' \citep{Kro20}.  Deciding whether a given comet is dynamically new or a returning LPC is usually not simple, however, because of astrometric uncertainties, and because hard-to-model non-gravitational forces due to anisotropic outgassing introduce uncertainty into the fitted orbits.  

Thermal and physical evolution of comets in both the Kuiper belt and Oort cloud reservoirs is slow,  but dramatic changes can occur on timescales of months, weeks, days and even hours when comets approach within a few astronomical units of the Sun.  These changes include simple sublimation losses but also nucleus breakup into resolved fragments and catastrophic disintegration leading to  complete cometary disappearance.  For example, in a recent study, 7/27 (i.e., $\sim$25\%) of well observed LPCs with perihelia $<$2 au did not survive perihelion \citep{Jew22}.  Dynamical models suggest that the probability of an LPC surviving $N$ perihelia is $\sim N^{-0.6}$ meaning, for example, that only half survive three successive perihelia \citep{Wie99}.  It is likely that the nucleus survival fraction, in addition to being dependent on the perihelion distance, is a function of nucleus size, with sub-kilometer LPC nuclei being more susceptible to breakup and disintegration than larger ones \citep{Jew22}.  

At present, the relative rates at which destructive processes occur, and the details of the underlying physical mechanisms, are poorly known.  In part, this is because useful observations are  particularly difficult to obtain when comets are near perihelion and appear projected at small angles from the Sun. Moreover, nucleus fragmentation and disintegration events are unpredictable and rapid, further limiting the opportunity to obtain diagnostic observations.  There is also a publication bias: all too often, comets that disintegrate or otherwise disappear when near perihelion simultaneously disappear from the minds of astronomers and from the scientific publication record.  For all these reasons, quantitative observations of comets close to the Sun are of special interest. 

Comet C/2021 O3 (Panstarrs) (hereafter ``O3'')  was discovered inbound at 4.31 au on UT 2021 July 26 by \cite{Wer21} about nine months before reaching perihelion at $q$ = 0.287 au on UT 2022 April 21. The osculating orbit is weakly hyperbolic, with semimajor axis $a$ = -1800 au, orbital eccentricity, $e$ = 1.00015, and inclination, $i$ = 56.8\degr.  The pre-entry barycentric orbital elements (corrected for planetary perturbations to 1900 January 1, when the heliocentric distance was 138 au) are  a = -9.86$\times10^4$ au, e = 1.000003, and i = 56°.6. As such, O3 qualifies as a dynamically new LPC and has probably made few, possibly zero, previous passes through the planetary region. Strong thermal effects on the nucleus of O3 are to be expected, given the high radiative equilibrium temperatures at perihelion (exceeding 700 K at the sub-solar point).  Indeed, comet O3 was reported to have disintegrated soon after perihelion (\cite{Zha22}, \cite{Com23}, \cite{Hol24}).  In this paper, we combine available observations to examine comet O3 before and after its encounter with the Sun.  

\section{OBSERVATIONS}

We utilize data from several sources in order to obtain a clear picture of the evolution of O3.   

\subsection{Cometas Optical Photometry:  }
\begin{figure}
\epsscale{0.7}

\plotone{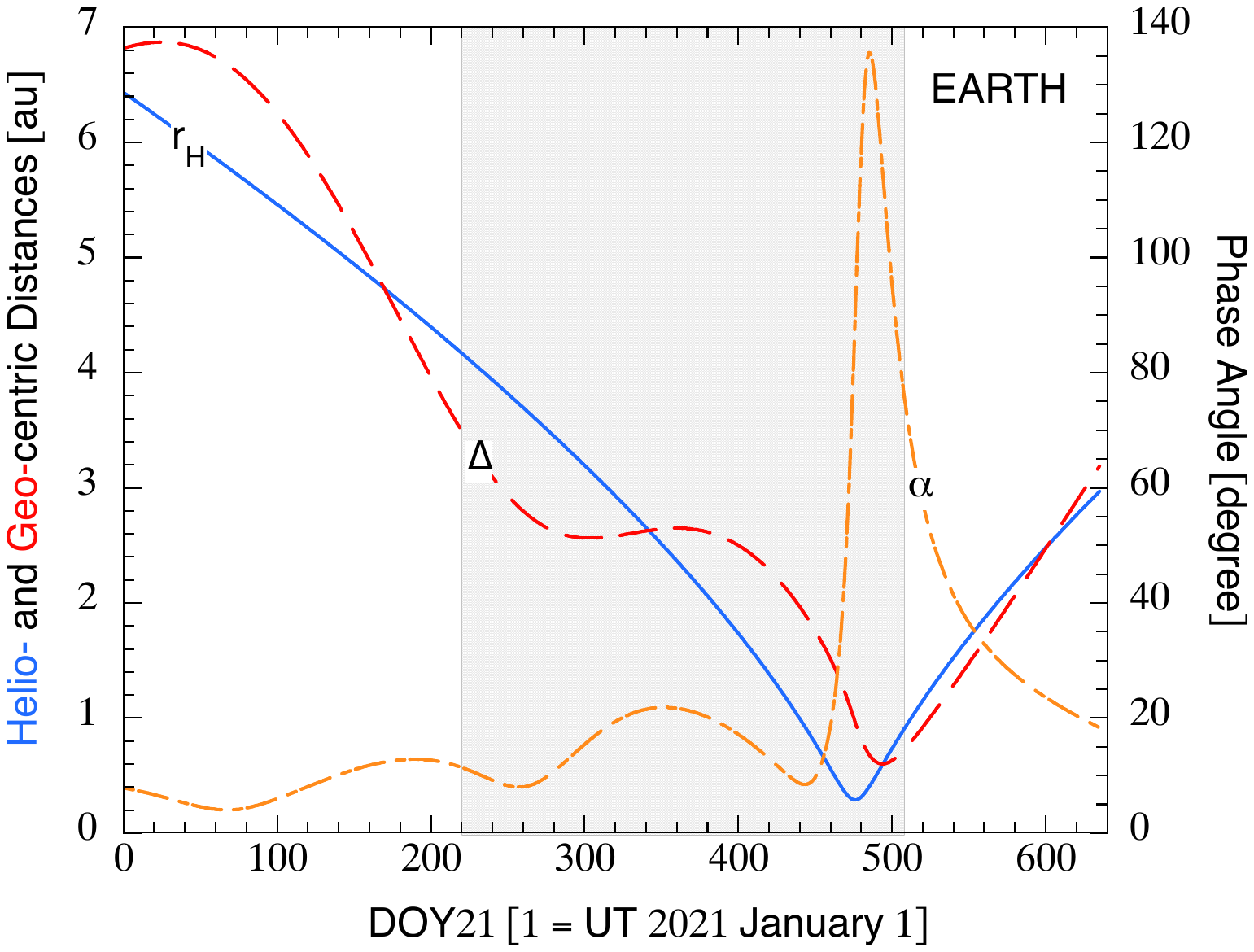}
\plotone{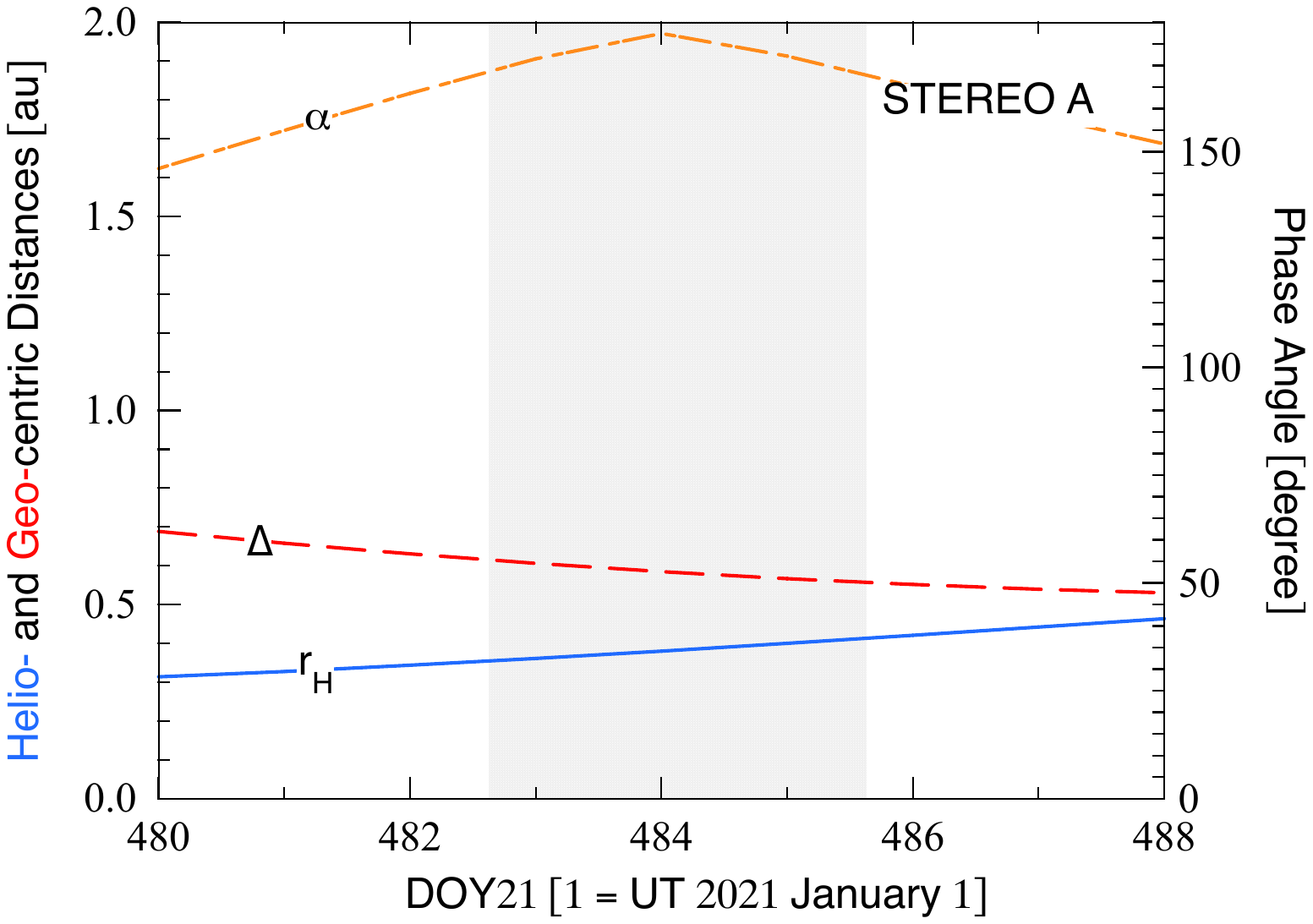}

\caption{(Upper:) Observing geometry as viewed from Earth as a function of observation date.  $r_H$ (solid blue curve) and $\Delta$ (dashed red curve) are the heliocentric and geocentric distances, respectively, while $\alpha$ (dashed-dot orange curve) is the phase angle.  The shaded region shows the range of dates on which observations were acquired.  (Lower:) Same quantities as in the upper panel but for observations from STEREO A.   \label{RDa}}
\end{figure}

We used optical observations obtained between UT 2021 August 8 ($r_H$ = 4.17 au inbound) through perihelion to 2022 May 20 ($r_H$ = 0.84 au outbound) as retrieved from the online Cometas Database\footnote{\url{http://www.astrosurf.com/cometas-obs/}}.  This compilation is particularly useful because the data  benefit from largely uniform observational and analysis techniques.  In contrast,  many reported measurements of cometary magnitudes are obtained using a variety of telescopes, filters and photometric extraction schemes, and consequently  suffer from   systematic and poorly defined errors.  The largest of these errors result from the use of widely different photometric apertures, which sample very different volumes of coma surrounding the nucleus.  In contrast, the Cometas measurements use a consistent set of apertures and are calibrated photometrically against the Gaia DR2(G) sky survey.  Here, we use the data obtained within a  10\arcsec$\times$10\arcsec~square aperture.  Measurements from larger apertures suffer increasingly from sky background uncertainties and field object contamination and are not considered here.  The Cometas photometry uses the Gaia G-band filter, which has 50\% peak transmission at $\sim$4000\AA~and 8500\AA, with a central wavelength near 6000\AA. The observing geometry for the Cometas data is shown in the upper panel of Figure \ref{RDa}, where the shaded region shows the range of dates over which observations were obtained.

\subsection{Archival Optical Photometry}

\begin{figure}
\epsscale{0.9}
\plotone{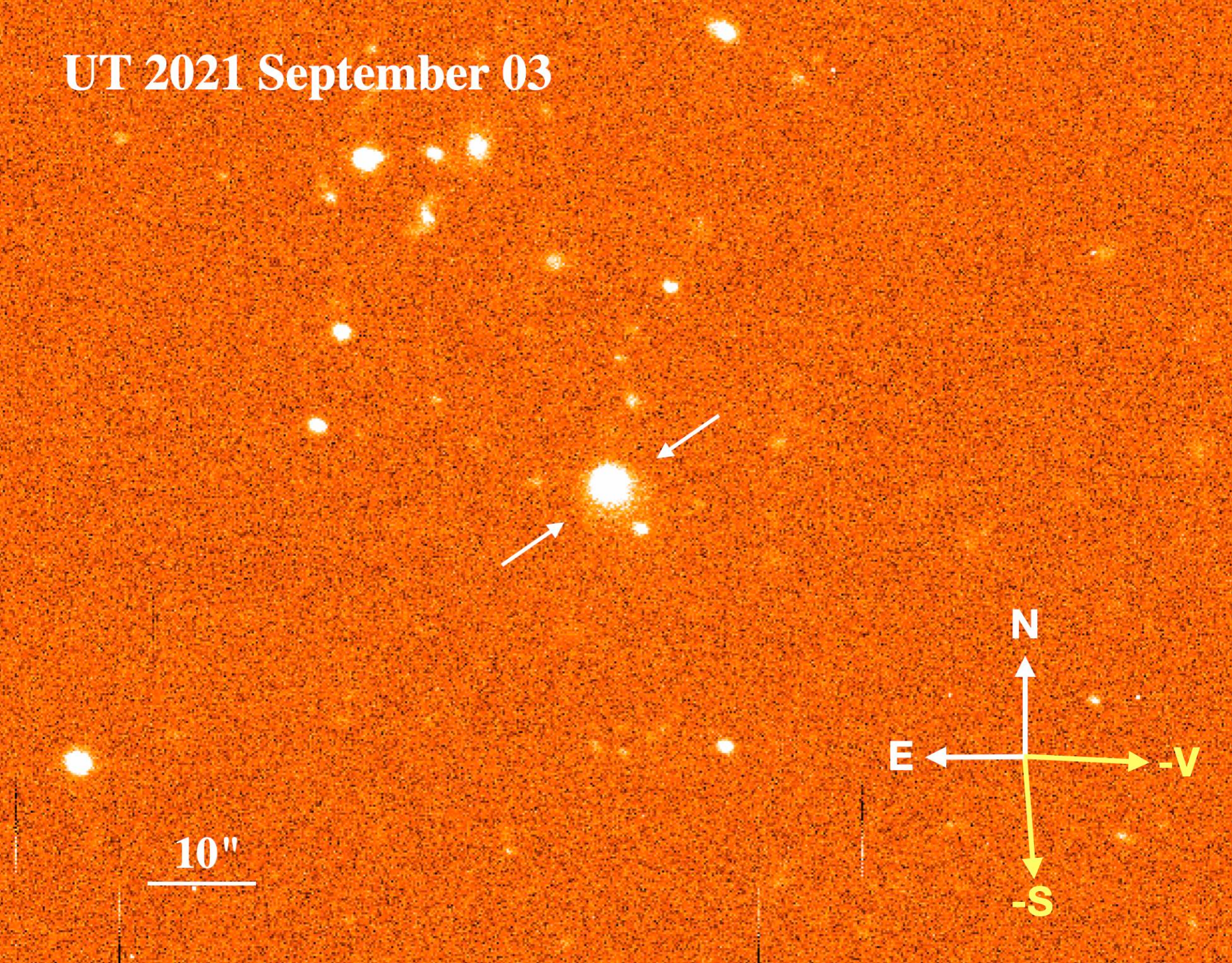}

\caption{Pre-perihelion appearance of C/2021 O3 on UT 2021 September 03 (DOY 246) at $r_H$ = 3.865 au.  This is a 60 s image through an r' filter from the CFHT 3.6 m telescope, archived at CADC.  A scale bar and cardinal directions are marked. \label{September03}}
\end{figure}

Figure \ref{September03} shows the appearance of the comet in images from the 3.6 m Canada France Hawaii Telescope (CFHT) taken UT 2021 September 03 when O3 was at $r_H$ = 3.865 au (example in Figure \ref{September03}).  Taken about six weeks after discovery, the comet is extended, with isophotes that are largely circular, and with only minimal evidence for a faint tail asymmetry towards the southwest (position angle $\sim$200\degr).

Observations from UT 2022 May 10 ($r_H$ = 0.62 au outbound) were downloaded from the CADC archive\footnote{\url{https://www.cadc-ccda.hia-iha.nrc-cnrc.gc.ca/en}}. These observations utilized the 0.8 m diameter Schmidt telescope of the Calar Alto Observatory in southern Spain using a ProLine PL230 16-bit CCD. The image scale was 1.29\arcsec~pixel$^{-1}$ giving a roughly 0.7\degr~field of view on the 2048$\times$2064 pixel detector. Observations were taken through the ``RG-clear'' filter, with which  the effective bandwidth is largely determined by the wavelength dependence of the CCD quantum efficiency (50\% peak efficiency at 3800\AA~to 9000\AA, center wavelength $\sim$6500\AA). Seeing during the measurements was approximately 3.\arcsec2 FWHM (full width at half maximum).

A set of sixty  exposures each of 20 s were obtained in the period UT 20:10 to 20:56, with on-sky dithering to provide protection against detector defects.  The first 26 images were found to possess a higher sky signal than the remaining 34, for unknown reasons.  We used only images taken between UT 20:31 and 20:56 (680 s total integration) for the analysis.   The Calar Alto data are limited by two effects.  First, the field star density on May 10 is very high (galactic latitude $\sim$0\degr) giving rise to star contamination problems in many images.  Second, the ProLine camera uses a thick silicon wafer that is susceptible to strong OH sky line interference fringes.  While these fringes were partially removed in image processing, they remain as a large source of error in the compiled data.

\subsection{Jaeger Observatory}
Post-perihelion observations from UT 2022 May 17 were acquired by coauthor Michael Jaeger using a 35 cm diameter, f/4.2 telescope located in Stixendorf, Austria and equipped with a back illuminated Sony CMOS detector cooled to -5\degr~C.  The detector, with 4788$\times$3194 pixels giving a field of view approximately 0.69\degr$\times$0.46\degr, was read out with the pixels binned 2$\times$2, resulting in an image scale 1.04\arcsec~pixel$^{-1}$.  Stellar images have 3.1\arcsec~FWHM. No filter was employed.  We assume that the wavelength passband approximates the quantum efficiency of the detector, which has half power points near wavelengths 3600\AA~and 7000\AA, and peaks near 5000\AA.

Twelve images each of 70 s integration (total time 840 s) were combined at sidereal rates and, separately, with offsets calculated to track the motion of the comet against the fixed stars (Figure \ref{May17}).  The sidereally stacked images were used to calibrate the data against field stars from the Gaia DR3 catalog\footnote{\url{https://www.cosmos.esa.int/web/gaia/dr3}}.  The post-perihelion comet shows a circular, centrally condensed morphology indicating the presence of an active source, with no evidence for a debris cloud or multiple components that might indicate nucleus breakup.


\subsection{STEREO-A COR2 Coronagraph:}
 The STEREO A spacecraft \citep{Eyl09} shares the orbit of the Earth but, at the time of the C/2021 O3 observations, was separated from our planet by $\sim$0.5 au.  STEREO observations therefore offer a different perspective on the comet, in addition to providing data at elongation angles smaller than can be reached from Earth-based observatories.  We used the COR2 coronagraphic imager \citep{How08},  to take $\sim$6000\AA~observations of O3 at extraordinarily small solar elongations, from the inner edge of the field of view at 2.5 solar radii ($\sim$1.25\degr), to the outer edge at  4\degr.   The COR2 image scale is 14.7\arcsec~per pixel. Stars measured in the COR2 data had FWHM = 3.0 pixels (44\arcsec).

      The observing geometry is shown in the lower panel of Figure \ref{RDa}.   We  observed O3 in COR2 between UT 2022 April 27.6 ($r_H$ = 0.35 au outbound) and April 30.6 ($r_H$ = 0.41 au outbound).  The comet was located between the Earth and the Sun, giving phase angles as seen from STEREO A in the range from 168\degr~to 177\degr, corresponding to a forward-scattering geometry. The elevation of the spacecraft above the orbital plane of O3 varied from +3.9\degr~to -0.2\degr, providing a nearly edge-wise view.  
  
     The principal impediment in using COR2 to study the comet is the structured, bright and variable background caused by the Sun's corona. Figure \ref{single_image} shows a single image in which the comet (arrows) is barely visible against the corona.  We suppressed the coronal background by median combining images in groups to remove time-variable background structures (Figure \ref{stereo2}).  The $\sim$3 day observing window was limited by the departure of O3 from the COR2 field of view.


\section{MEASUREMENTS}

Casual inspection of the data shows that O3 survived perihelion.   In both pre-perihelion and post-perihelion ground-based observations, O3 stands out as a clearly resolved, centrally condensed active comet with a circularly symmetric coma, as shown in Figures \ref{September03} and \ref{May17}.  In near perihelion, forward-scattering observations from STEREO the comet appears as a linear debris sheet, with no evidence for a nucleus (Figures \ref{single_image} and \ref{stereo2}).  These morphological changes are consistent with the different observing geometries from the Earth and STEREO, as described below.

We measured the apparent magnitudes of O3 in the ground-based data using circular photometry apertures of projected radius 10\arcsec. This aperture was selected in order to be consistent with the measurements from the Cometas database and to minimize the effects of sky background uncertainty, which grow with the aperture radius.  Sky subtraction was achieved using the median signal within concentric annuli whose inner and outer radii were typically 50\arcsec~to 80\arcsec~(selected on the basis of the image quality and the density of field objects).  Photometric calibration used field stars and data from the Sloan DSS and the Pan STARRS sky surveys.  The different instruments used to measure the comet employed different, generally non-standard, extremely broad filters with central wavelengths in the V to R region of the spectrum.  Lacking information about the optical color of O3, we did not attempt to color correct the photometry, other than by using reference stars from Sloan DSS and Pan STARRS having near solar colors.  Evidence that color corrections are small ($\lesssim$0.2 magnitude) is provided by the consistency of measurements at a given epoch, from different telescopes, evident in Table \ref{geometry}.

\subsection{Heliocentric Lightcurve:} 
\label{section3.1}
We used the ground-based photometry to determine the heliocentric lightcurve of O3.  To remove geometric effects affecting the apparent brightness, we employ the inverse square relation written as

\begin{equation}
p\Phi(\alpha) C(r_H) = 2.25\times10^{22}\pi r_H^2 \Delta^n 10^{-0.4(V - V_{\odot})}
\label{inversesquare}
\end{equation}

\noindent where $p$ and $C(r_H)$ are the albedo and scattering cross-section within the photometry aperture, respectively, $V$ and $V_{\odot}$ = -26.73 are the apparent V magnitudes of the comet and the Sun, respectively,  the heliocentric and geocentric distances are $r_H$ and $\Delta$, respectively, both in au, and $n$ is a constant.   Quantity $\Phi(\alpha)$ represents the phase function, falling from unity at phase angle $\alpha$ = 0\degr~to small values at larger phase angles but eventually rising above unity in the forward scattering direction.  We used the generic cometary phase function tabulated by Schleicher\footnote{\url{https://asteroid.lowell.edu/comet/dustphase/}}, while recognizing that this function is particularly uncertain at the largest phase angles.  For a target subtending an angle small compared to the photometry aperture, $C$ is independent of $\Delta$ and then $n$ = 2, to conform to inverse-square behavior.  However, for a cometary coma larger than a fixed  photometry aperture, the cross-section increases with $\Delta$, because at larger distances an aperture of fixed angular size samples a larger volume of coma, a fact known as the ``$\Delta$ effect'' \citep{Mar86}.  For a steady-state coma, the integrated cross-section within a fixed aperture scales in proportion to $\Delta$, which corresponds to $n$ = 1.  We assume this value for O3 in Equation \ref{inversesquare} while recognizing that O3 is a dynamic body and there is no physical reason why a fixed value of $n$ should prevail at all distances and times.  We write

\begin{equation}
C(r_H) = C_1 r_H^{-s}
\label{steady}
\end{equation}

\noindent where $C_1$ is the cross-section in the photometry aperture at $r_H = \Delta$ = 1 au and $s$ is the heliocentric index.  Substituting  Equation \ref{steady} into \ref{inversesquare} and rearranging, we obtain

\begin{equation}
V = V_{\odot} - 2.5\log\left(\frac{p \Phi(\alpha) C_1  }{2.25\times10^{22}\pi \Delta r_H^{\eta} }  \right)
\end{equation}

\noindent where $\eta = s + 2$.  Equivalently, we write


\begin{equation}
V = H + 2.5 \eta \log(r_H) + 2.5\log(\Delta) -2.5\log(\Phi(\alpha)),
\label{fit}
\end{equation}

\noindent where 

\begin{equation}
H = V_{\odot} - 2.5\log \left(\frac{p C_1}{2.25\times10^{22} \pi} \right)
\end{equation}

\noindent is the absolute magnitude (the apparent magnitude at $r_H = \Delta$ = 1 au and $\alpha$ = 0\degr).    
For an asteroid or inert comet nucleus, the scattering cross-section is independent of $r_H$ and so $s$ = 0 by Equation \ref{steady}, leaving $\eta$ = 2. For a sublimating body, the cross-section in the coma increases as the comet becomes more active towards decreasing $r_H$, giving $s >$ 0 and $\eta > 2$.  Close to the Sun, in the limiting case where all the absorbed energy is used to sublimate ice, energy balance gives $s$ = 2 and so $\eta$ = 4.  For this reason, the apparent magnitudes of comets are often extrapolated from data assuming  2.5$\eta$ = 10 and, by Equation \ref{fit},  the absolute magnitude is then referred to as H$_{10}$. 


A least-squares fit of Equation \ref{fit} to the pre-perihelion 10\arcsec~aperture photometry of O3 gives $\eta = 2.59\pm$0.21
, from 3.8 au to 1.9 au, indicating a very weak ($C \propto r_H^{-0.59\pm0.21}$)  rise in dust production with decreasing $r_H$ (Figure \ref{Cometas_plot}).  Similarly, \cite{Hol24} found $\eta = 2.5\pm$0.5  ($C \propto r_H^{-0.5\pm0.2}$), over the 3.6 $\le r_H \le$ 3.0 au distance range while \cite{Lac25} found $\eta = 2.4\pm0.1$ ($C \propto r_H^{-0.4\pm0.1}$), over  2.2 $\le r_H \le$ 2.9 au. The small empirical values of $\eta$, all in mutual agreement, explain why early predictions of extraordinary perihelion brightness of O3, based on an assumed  $\eta$ = 4, were not met.  The fit also gives $H  = 13.00\pm0.26$. Substitution into Equation \ref{inversesquare} gives $C_1 =$ 225 km$^2$ for an assumed $p_V$ = 0.04.  If contained entirely within a compact nucleus, this cross-section would equal that of a sphere $a = (C/\pi)^{1/2} \sim$ 8.5 km in radius.  However, comet O3 was clearly cometary in nature even in the earliest observations at the largest heliocentric distances, and so we interpret $a$ = 8.5 km as setting a strong upper limit to the pre-perihelion nucleus radius.  

 After perihelion, the comet appeared $\sim$3.5 magnitudes fainter than  expected from the extrapolated best-fit pre-perihelion model (see dashed curve in Figure \ref{Cometas_plot}), indicating a loss of cross-section in the coma, at any given heliocentric distance, by a factor of $\sim$25. The post-perihelion absolute magnitude, $H$ = 16.5, gives $C$ = 9 km$^2$ for the cross-section within the 10\arcsec~aperture, and the corresponding upper limit to the post-perihelion  nucleus radius is $a$ = 1.7 km, again assuming geometric albedo $p$ = 0.04. 

\subsection{STEREO Photometry Near Perihelion:} 

A composite formed from 215 COR2 images is shown in Figure \ref{stereo2}, showing a morphology quite distinct from that  in pre- and post-perihelion ground-based data.  The appearance is that of a linear dust trail, aligned approximately with the projection of the orbit into the plane of the sky.  There is no evidence for a head or nucleus in the COR2 data but the point-source sensitivity is limited (the limiting magnitude is position dependent but field stars fainter than V $\sim$ 13 cannot be reliably detected in star-aligned image composites).  As mentioned above, the COR2 images show the comet in a forward scattering geometry, and the phase function of cometary dust, although rarely measured at the extreme angles of O3, is generally forward scattering \citep{Mar86}. 

The COR2 STEREO data present many challenges to accurate photometry, given the spatial complexity and temporal variability of the sky background, the large (14.7\arcsec) pixels, and the relative faintness of O3 (c.f.~Figure \ref{single_image}).  Nevertheless, we attempted to measure the apparent brightness of the trail in the COR2 data.  To calibrate the data, we identified a set of field stars having known magnitudes and spectral types from the SIMBAD website\footnote{\url{https://simbad.cds.unistra.fr/simbad/sim-fid}}.  We selected stars of spectral types F, G and K in order to approximate the color of the Sun, and measured their fluxes within an aperture 5 pixels (73.5\arcsec) in radius, with sky subtraction from the median signal within a contiguous annulus having outer radius 20 pixels (294\arcsec).  The measurements show a scatter of $\pm$0.5 magnitudes that we attribute to residual systematic variations in the background owing to variable coronal structures.  

To capture the light from O3, we used rectangular regions aligned with the dust trail.  Again because of structure in the sky background, we could not reliably measure the whole visible trail.  Using a box 450 pixels (6615\arcsec) in length and 17 pixels (250\arcsec) wide, with sky level determined in adjacent boxes of the same size, we obtained an integrated magnitude $V = 5.4\pm$1.0, where we have generously estimated the uncertainty based on experimentation with different apertures.  This photometry box excludes the south-eastern end of the trail, where background removal is particularly problematic, and therefore gives a lower limit to the brightness.  The average phase angle of the COR2 composite image is $\bar{\alpha}$ =  171.2\degr~and it is likely that the enhanced brightness is in part a result of strong forward scattering.  Using the aforementioned Schleicher phase function, for which $\Phi(\alpha) \sim$ 220, we find a scattering cross-section $C \sim$ 56 km$^2$, intermediate between the pre-perihelion and post-perihelion values determined from ground-based photometry.  

We carefully measured the position angle of the trail, $\theta_{PA}$, using a series of surface brightness profiles spaced along its length, finding $\theta_{PA}$ = 123.9\degr$\pm$0.5\degr.  To help interpret this position angle, we next computed a set of synchrones for particles released 10, 20, 40, 80, 160 days prior to the observation date and syndynes for particles with $\beta$ = 1, 0.3, 0.1, 0.03, 0.01 (i.e.~nominal particle radii 1, 3, 10, 30 and 100 $\mu$m, respectively) for comparison with the STEREO image (Figure \ref{curves}).  Figure \ref{synsynd} shows the model position angles as functions of $\beta$, in the case of the syndyne models (solid red curve), and the particle release times (relative to the date of observation) for the synchrones (dashed blue line).  The measured position angle is shown as a horizontal yellow band.  The intersections of the models with the yellow band give the allowable range of parameters for the syndyne and synchrone representations.

The position angles of the synchrone (particles having a range of sizes released at one time) and syndyne (particles of one size released continuously) models both occupy a narrow range (from 122\degr~to 127\degr) because the STEREO observatory is only a few degrees above the orbital plane of O3 and projection effects are dominant.  Consequently, we cannot reliably use the position angle to discriminate between the synchrone and syndyne approximations.  If interpreted as a pure synchrone, Figure \ref{synsynd} (dashed blue curve) shows that the measured $\theta_{PA}$ is best matched by ejection dates 43$_{-17}^{+44}$ days prior to the UT 2022 April 29 observation (corresponding to ejection dates from 2022 February 1 - April 3).  On the other hand, if interpreted as a pure syndyne, the measured $\theta_{PA}$ indicates best-fit values $\beta = 0.11_{-0.03}^{+0.04}$, corresponding to particle radii from $a \sim$ 7 to 13 $\mu$m.  Note that substantially smaller ($\beta$ = 1, $a$ = 1 $\mu$m) and larger mean particle sizes (e.g.~$\beta$ = 0.05, $a$ = 20 $\mu$m) are ruled out by the plot.  We take the nominal particle radius from the syndyne plot as $a \sim$ 10 $\mu$m.

The forward scattering geometry provides an independent measure of the particle size, because diffraction around a particle of radius $a$ forward scatters into an angle  $\psi \sim \lambda/(2a)$. For example, the smallest  STEREO scattering angle was $\psi = (180 - \theta) \sim$ 2.5\degr~(Table \ref{geometry}).   Expressing $\psi$ in radians,  the scattering relation gives $a \sim$ 7 $\mu$m as the size of the particles most likely dominating the STEREO signal, compatible with the size of particles deduced from the syndyne model.

\subsection{SOHO/SWAN Lyman-$\alpha$ Observations:}
Lyman-$\alpha$ observations from the SOHO/SWAN ultraviolet imager provide a measure of the cometary water production rate, $Q_{H_2O}$ \citep{Com23}.  The water production rate near 0.5 au pre-perihelion was $Q_{H_2O} = (5\pm2)\times10^{28}$ s$^{-1}$ (1500$\pm$600 kg s$^{-1}$) which, scaled to 1 au by the inverse square law, gives  $Q_{H_2O} = (1.2\pm0.5)\times10^{28}$ s$^{-1}$ ($\sim$375 kg s$^{-1}$).  The Lyman-$\alpha$ measurements show a rise from about 13 to 10 days pre-perihelion, reaching a peak  $Q_{H_2O} = (9.3\pm2.5)\times10^{28}$ s$^{-1}$ (2800$\pm$750 kg s$^{-1}$) on UT 2022 April 11 (when $r_H$ = 0.42 au).  However, in data taken after April 14 (7 days pre-perihelion) the uncertainty on $Q_{H_2O}$ exceeds its measured value and the observations become unconstraining.  At  0.42 au, a perfectly absorbing spherical water ice nucleus would sublimate from the Sun-facing hemisphere in equilibrium  with sunlight at the rate $f_s = 1\times10^{-3}$ kg m$^{-2}$ s$^{-1}$.  If in steady-state, the peak production would therefore correspond to an exposed ice area $C_i = \mu m_H Q_{H_2O}/f_s = (2.8\pm0.8)\times10^6$ m$^2$ (2.8$\pm$0.8 km$^2$).   This is $C_i/C_1 \sim$ 1\% of the 225 km$^2$ scattering cross-section inferred from $H$, showing that ice is a negligible component of the cometary dust grains interior to 1 au.   We estimate the rate of change of radius of a sublimating ice grain from $da/dt \sim f_s/\rho$ which, with density $\rho$ = 1000 kg m$^{-3}$, is $da/dt \sim$ 1 $\mu$m s$^{-1}$. A 10 $\mu$m ice grain would last only $\sim$10 s, explaining the non-icy nature of the dust in O3 near perihelion.


\section{DISCUSSION}

\begin{figure}
\epsscale{0.95}
\plotone{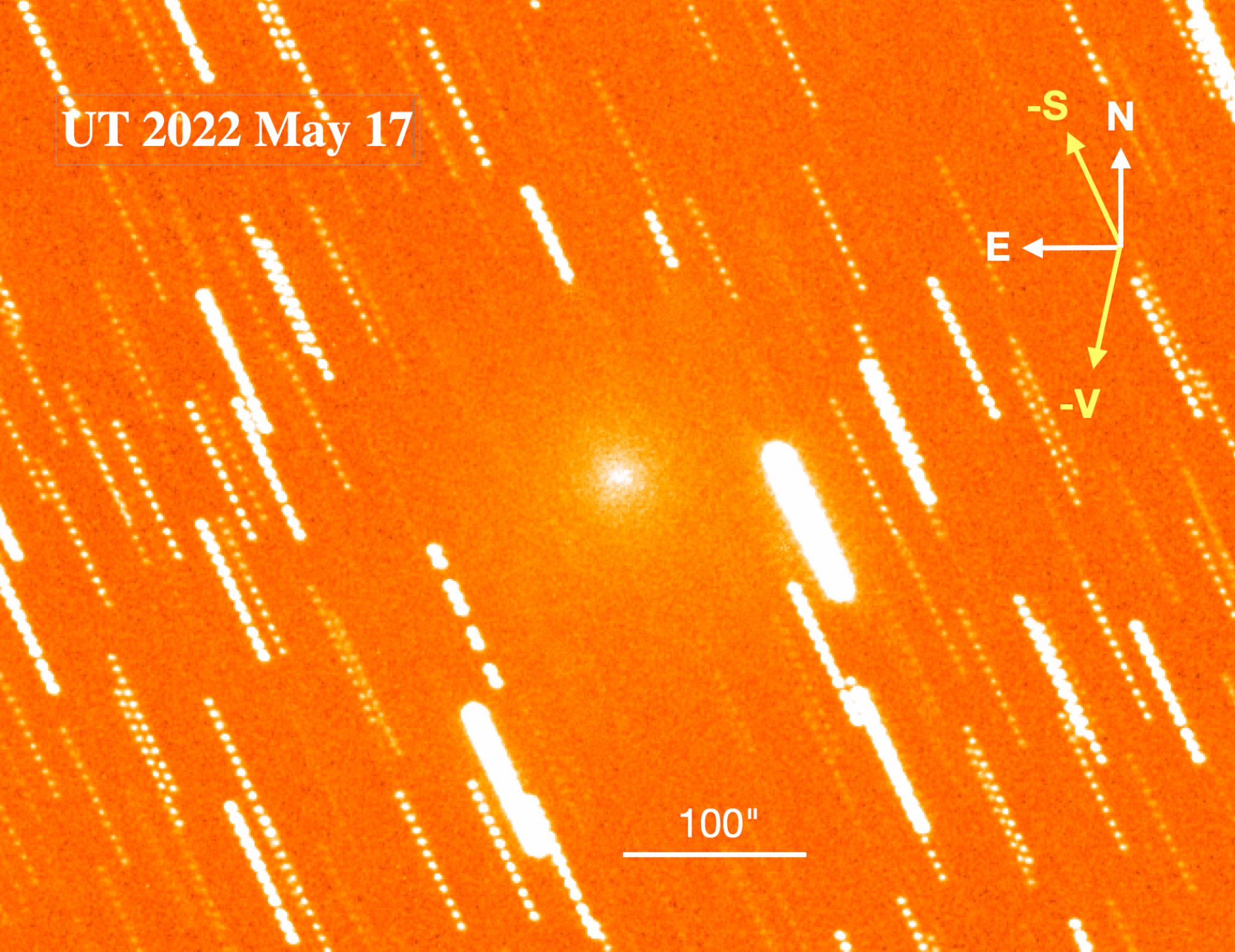}

\caption{Post-perihelion appearance of C/2021 O3  on UT 2022 May 17 (DOY 502), when $r_H$ = 0.80 au, with cardinal directions and a 100\arcsec~scale bar.  This is a cleaned composite of twelve images selected to avoid contaminating stars. \label{May17}}
\end{figure}

\begin{figure}
\epsscale{0.95}
\plotone{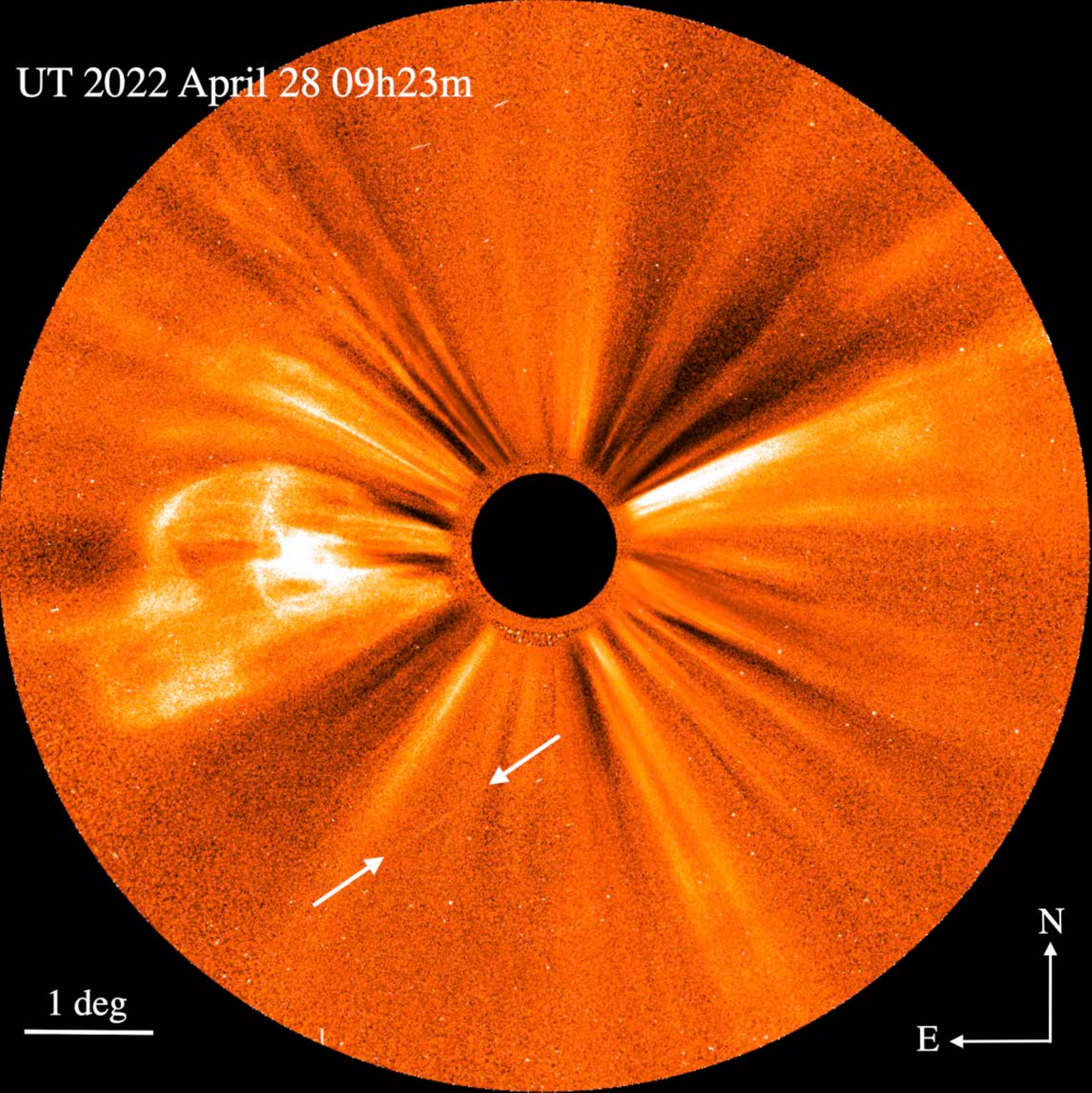}

\caption{Single COR2 image at $r_H$ = 0.37 au (DOY 483 - 484) showing the full 8.4\degr~wide field of view, the central occulting spot, and coronal structures.  C/2021 O3 is marked between the two arrows. The cardinal directions and a 1\degr~scale bar are also shown. \label{single_image}}
\end{figure}

\begin{figure}
\epsscale{0.95}
\plotone{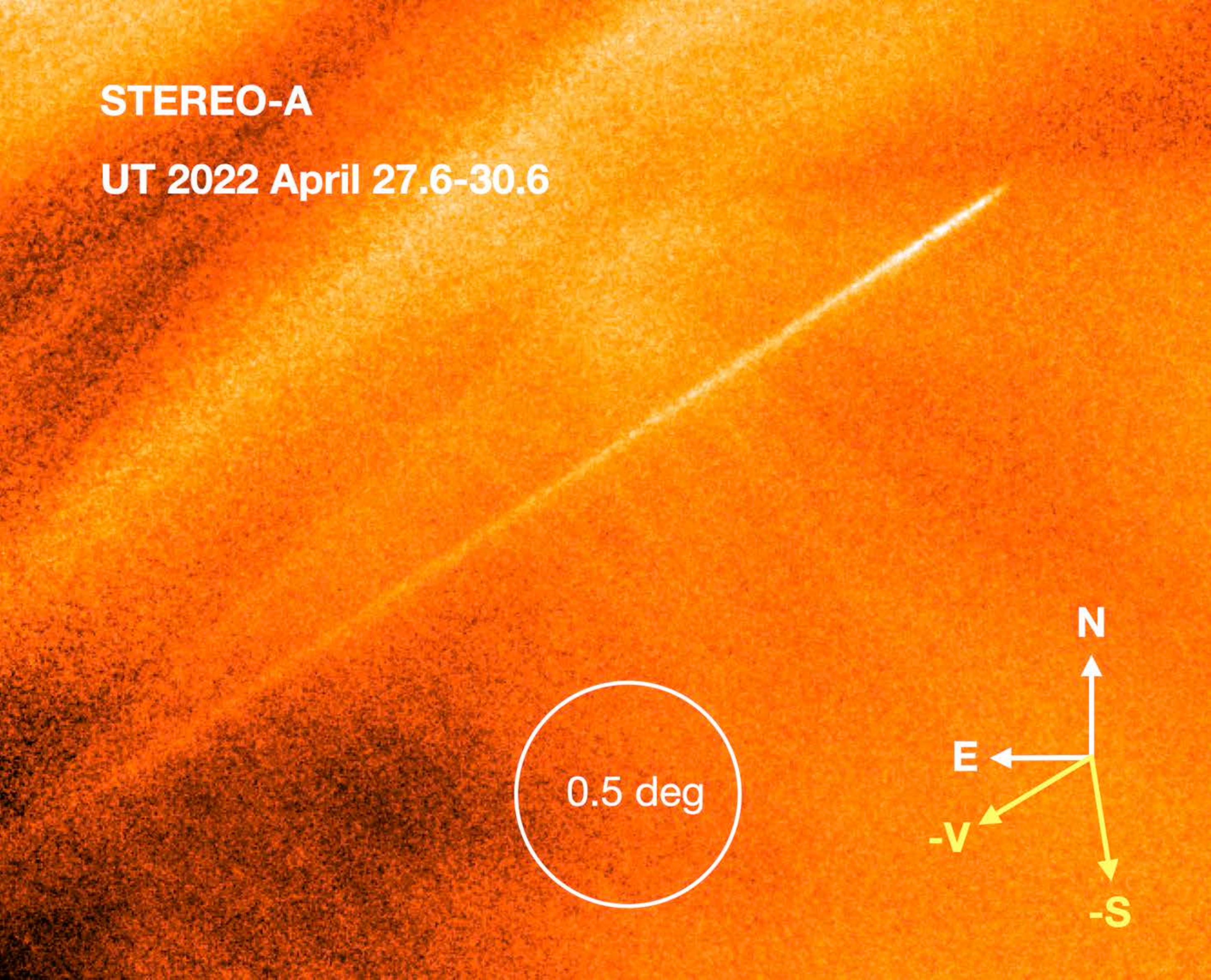}

\caption{Composite from STEREO-A computed from 173 COR2 images taken in the period  UT 2022 April 27.6 to 30.6 (DOY 482 to 485). Arrows show the projected negative heliocentric velocity vector ($-V$) and the projected anti-solar direction ($-S$), both for the mid-time (UT 2022 April 29.1) of the observation sequence.  The cardinal directions and a 0.5\degr~diameter circle representing the apparent diameter of the Sun are also shown. \label{stereo2}}
\end{figure}

\begin{figure}
\epsscale{0.8}
\plotone{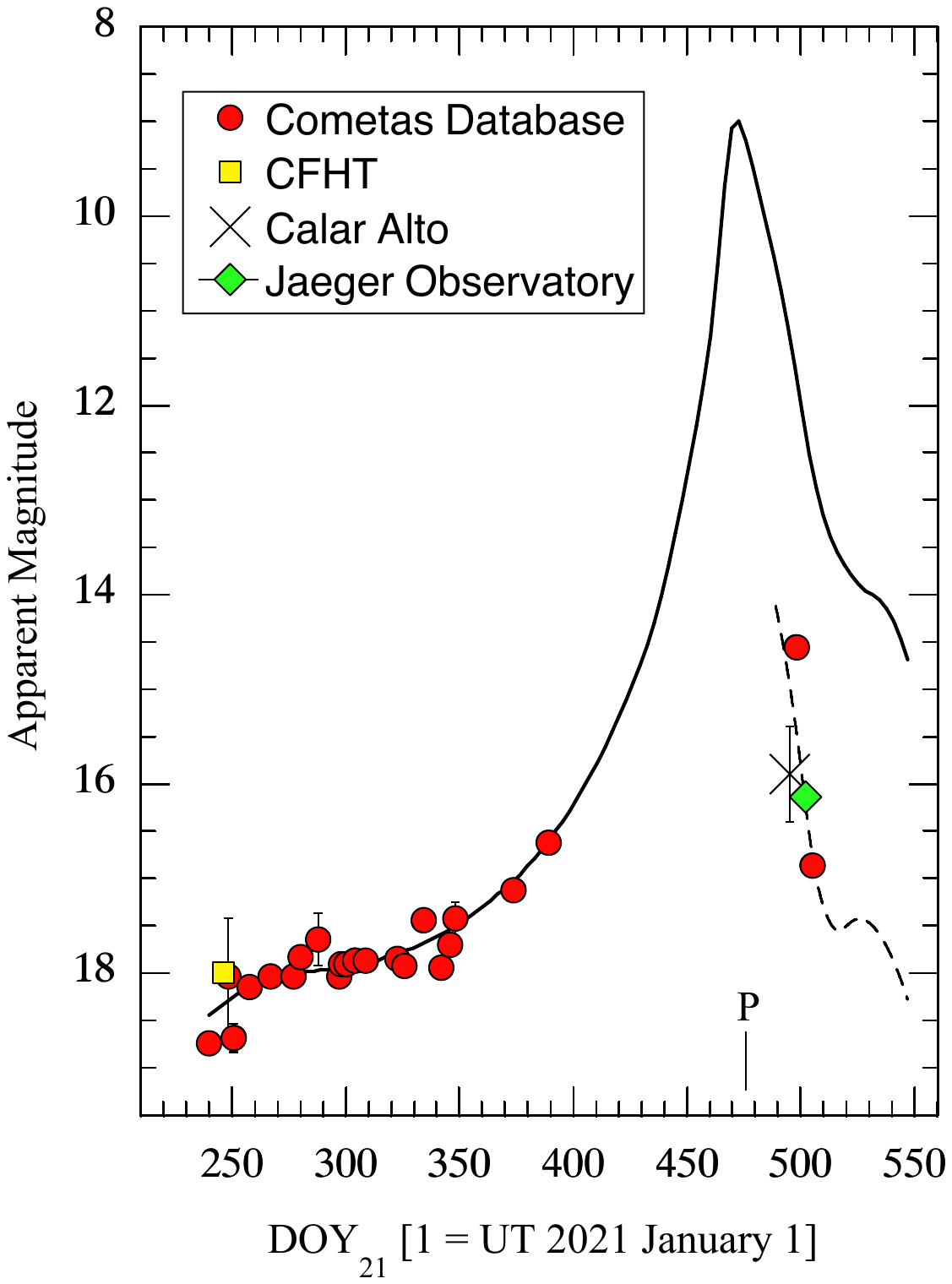}

\caption{Photometry within 10\arcsec~apertures extracted from images in the Cometas archive (filled red circles) and from CFHT (yellow square), Calar Alto (X) and Jaeger Observatory (green diamond).  The solid curve shows Equation \ref{fit} with $H$ = 13.00$\pm$0.26 and $n$ = 2.59$\pm$0.21 (solid black line). The dashed curve shows the same model but displaced by +3.5 magnitudes to better match the post-perihelion data. The date of perihelion is marked P.  \label{Cometas_plot}}
\end{figure}

\begin{figure}
\epsscale{0.95}
\plotone{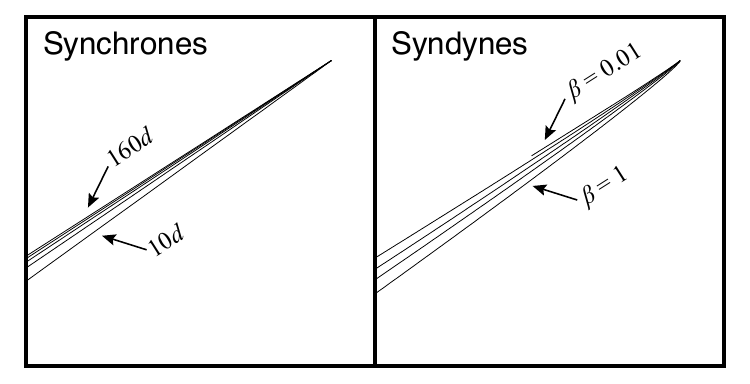}
\caption{(Left:) Synchrones for ejection 10, 20, 40, 80, 160 days prior to the April 29 (DOY 484) observation date and (right:) syndynes for particles with $\beta$ = 1, 0.3, 0.1, 0.03, 0.01 (i.e., nominal particle radii 1, 3, 10, 30 and 100 $\mu$m, respectively). Both panels show a region 4\degr$\times$4\degr~in size. \label{curves}}
\end{figure}

\begin{figure}
\epsscale{0.99}
\plotone{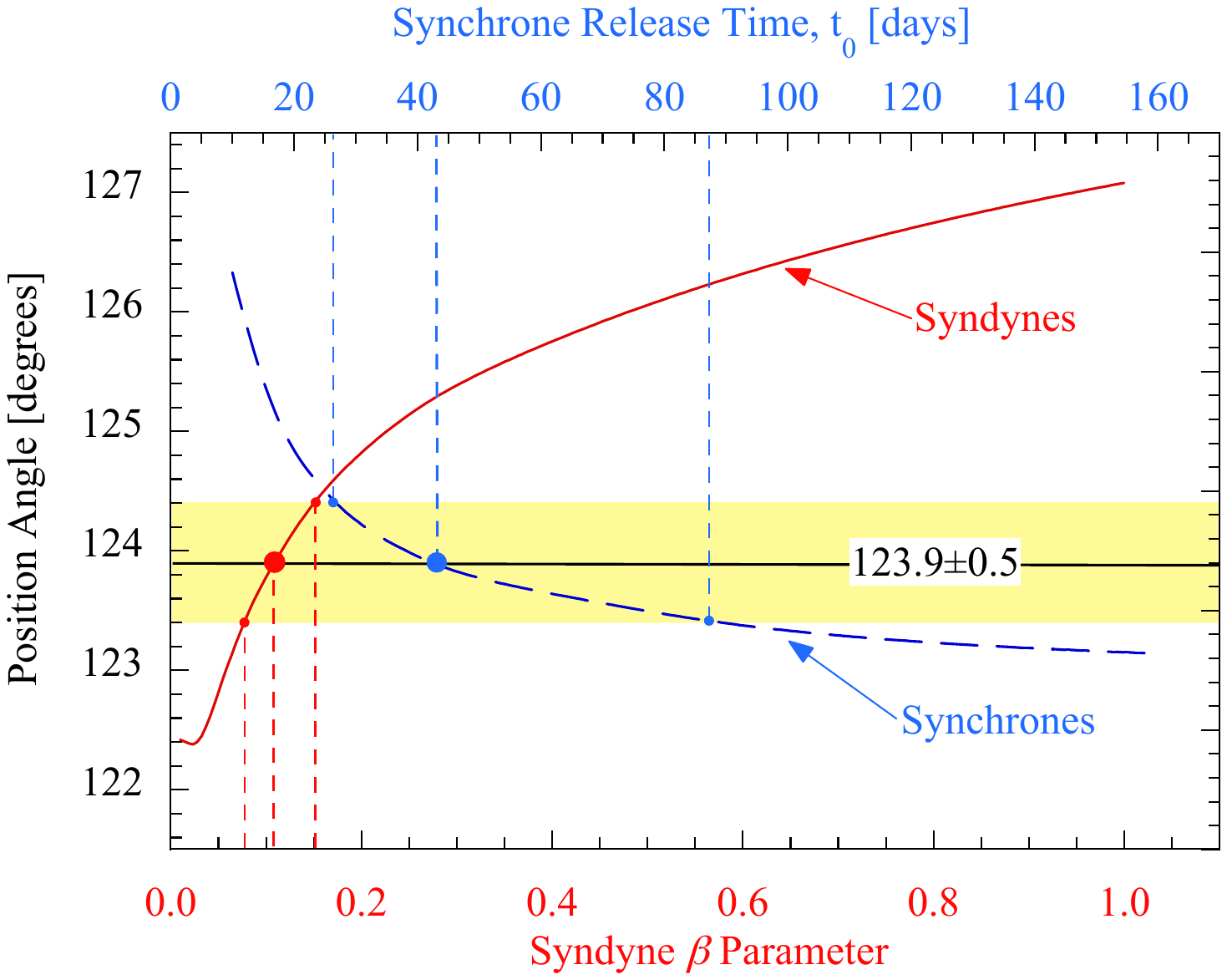}

\caption{Position angle expected from syndyne (solid red line) and synchrone (dashed blue line) models as a function of particle release date (measured in days prior to April 29 (DOY 484)) and $\beta$, respectively.  The horizontal black line and yellow band mark the measured position angle and $\pm$1$\sigma$ uncertainty.  The derived estimates of release time and $\beta$ are marked with blue and red vertical lines, respectively. \label{synsynd}}
\end{figure}

\textbf{Reported Non-Detection:} \cite{Zha22} failed to detect O3 on UT 2022 April 29 (DOY$_{21}$ = 484, or 8 days after perihelion) down to limiting magnitude $r' >$ 14 and concluded that the nucleus had not survived perihelion.   However, we clearly detected O3 from STEREO on the same date, and on later dates in post-perihelion ground-based data  (Table \ref{geometry}, see also Figure \ref{May17}).  Our observations indicate that O3 was simply too faint to be detected in the April 29 observation by \cite{Zha22}.  These authors did report a diffuse, co-moving, 40\arcsec~diameter structure 2\arcmin~southeast of the ephemeris location, which they interpreted as the possible decay product from the disintegration of O3. We find no evidence for such an offset, comoving structure in our data.

\textbf{Small Heliocentric Index:} As noted above, the measured scattering cross-section varies as $C \propto r_H^{-0.59\pm0.21}$  over the distance range 3.9 au to 1.9 au.  In detail, the heliocentric variation of the outgassing is a function of numerous unmeasured parameters both on the surface of the nucleus (size, shape, spin, obliquity and spatial distribution of ice)  and beneath it (thermal diffusivity of the bulk cometary material, aggregate structure of the interior) (c.f., \cite{Pri04}).  To a first approximation, however, the scattering cross-section can be assumed to be proportional to the instantaneous production rate from the nucleus.  In the 
strong sublimation limit, where all the energy absorbed from the Sun is used to break molecular bonds in the sublimation of ice, the sublimation rate should vary as $f_s \propto r_H^{-2}$.  This is the case for highly volatile carbon monoxide and carbon dioxide ices across the 2 to 4 au range.  Less volatile water ice follows a steeper variation with $r_H$, $f_s \propto r_H^{-4}$ to $r_H^{-8}$, becoming progressively steeper as $r_H$ increases and the fraction of absorbed energy used to sustain thermal radiation $\rightarrow$ 1. A simple comparison with the data then shows that water ice, although it is presumably responsible for the release of hydrogen giving the  Lyman-$\alpha$ signal \citep{Com23}, is not a plausible driver of activity in O3, and that the supervolatiles offer a better but still imperfect fit to the heliocentric variation.  

The discrepancy between the expected $r_H^{-2}$ and the observed $r_H^{-0.59\pm0.21}$ variation remains significant.  This difference is at least partly a result of the use of a fixed angular aperture for the photometry (c.f.~\cite{Jew21}).  To see this, note that the instantaneous cross-section of the dust within the aperture is roughly $C = (dC/dt) \tau_c$, where $dC/dt$ [m$^2$ s$^{-1}$] is the rate of production of dust cross-section (assumed proportional to the sublimation rate) and $\tau_c$ is the residence time of the dust in the aperture.  We approximate the latter by $\tau_c = \phi S \Delta/V$, where $\phi$ [arcsecond] is the angular radius of the photometry aperture, $S = 7.25\times10^5$ [m (arcsecond)$^{-1}$] is the length scale subtending 1\arcsec~at 1 au, and $\Delta$ is the geocentric distance in au.  If dust speed $V$ is independent of heliocentric distance, then we expect $C \propto (dC/dt) \Delta$ and, with $\Delta \sim r_H$ (see Figure \ref{RDa}) an intrinsic variation in the production rate, $dC/dt \propto r_H^{-2}$, leads to a more gentle variation $C \propto r_H^{-2} r_H \propto r_H^{-1}$ that is within 2$\sigma$ of the measured value $C \propto r_H^{-0.59\pm0.21}$.  A slight heliocentric dependence of the dust bulk speed would bring more perfect agreement (for example, $V \propto r_H^{-1/2}$ is commonly adopted in Monte Carlo models of comets (e.g., \cite{Mor22}) and would give $C \propto r_H^{-1/2}$, close to the observed variation).  Similar arguments were advanced to explain measurements of C/2017 K2, where $C \propto r_H^{-1.14\pm0.05}$ in photometry with a fixed linear (as opposed to angular) aperture \citep{Jew21}.  Lightcurve studies of other  LPCs show that small heliocentric indices are common (\cite{Hol24}, \cite{Lac25}), presumably at least in part because of the heliocentric variation of the residence time.

\textbf{Dust Mass:} In an optically thin collection of spherical particles, the dust mass is related to the cross-section by 

\begin{equation}
M = 4/3 \rho \bar{a} C
\end{equation}

\noindent where $\bar{a}$ is the average particle radius.  Assuming nominal $\rho$ = 10$^3$ kg m$^{-3}$ and using $C$ = 225 km$^2$ from the lightcurve model fitted to the pre-perihelion photometry,  we find $M = 3\times10^5 (\bar{a}/\textrm{1~}\mu\textrm{m}$) [kg].  With mean particle radius  $\overline{a}$ = 10 $\mu$m, as suggested by  the syndyne/synchrone analysis and forward scattering, we obtain dust mass $M \sim 3\times10^6$ kg, but this is clearly no better than an order of magnitude estimate.  Aperture residence times assuming $V$ = 500 m s$^{-1}$ at 1 au are $\tau_r \sim$ 8 hours (2.9$\times10^4$ s), giving a crude mass production rate $dM/dt = M/\tau_c \sim 10^2$ kg s$^{-1}$. This is of the same order as the 300 kg s$^{-1}$ water production rate inferred from Lyman-$\alpha$ observations scaled by the inverse square law to the same distance \citep{Com23}.

\textbf{Processes:} As noted above, O3 brightened only slowly on approach to the Sun and was fainter by a factor $\sim$25 after perihelion relative to the best-fit pre-perihelion model. Although O3 did not disappear, the overall impression is of a fading comet whose dwindling mass loss rates were unable to keep pace with the rapidly rising insolation\footnote{As an aside, we note that Evangelista-Santana et al. (2023) reported that O3 faded
(i.e., s $<$ 0) in pre-perihelion observations between 3.3 au and 2.3 au. While neither our observations nor those of \cite{Hol24} or \cite{Lac25}  support this claim, they agree in showing a distinctively slow change in brightness with decreasing $r_H$.}.  We consider possible causes of this progressive fading.  

1) The most dramatic interpretation of the pre- vs.~post-perihelion lightcurve asymmetry (Figure \ref{Cometas_plot}) would be that the nucleus was partially destroyed near perihelion by a catastrophic breakup. While this cannot be ruled out, an empirical argument against this possibility is that the scattering cross-section in near-perihelion STEREO data ($C \gtrsim$ 56 km$^2$) was less than in the pre-perihelion ground-based data (225 km$^2$) but more than in the post-perihelion data (9 km$^2$).  The variation of the cross-section is therefore more consistent with a steady decline that spans our data than with a surge near perihelion that would be expected of catastrophic breakup.  Separately, the circularly symmetric post-perihelion coma (Figure \ref{May17}) indicates the persistence of a single central source, and provides no evidence for the more diffuse, sometimes multiple component comae observed in clearly disintegrated comets. 

2) It is also unlikely that the fading results from simple nucleus shrinkage as a result of accumulated sublimation losses upon approach to the Sun. The loss of ice on the way to perihelion is energetically limited to a layer thickness only $\Delta Z \sim$ 4 m (Figure 6 of \cite{Jew25}).  This thickness is very small compared to any plausible estimate of the radius of the nucleus of O3 and so could not materially affect its sublimation rate through nucleus shrinkage.

3) Fading could result if the nucleus has a shell structure, with volatile ices confined to a surface layer of depth $\sim \Delta Z$, and with more refractory material (silicates, organic particles?) beneath.  This possibility would echo the suggestion of \cite{Oor51} to the effect that LPC nuclei may be covered in a frosting of super-volatile ice which is depleted after a few orbits.  We cannot rule out this possibility, although its success would depend on an unexplained and rather ad-hoc coincidence between $\Delta Z$ and the thickness of the ice shell.  

4) The most prosaic but nevertheless plausible interpretation is that the fading reflects seasonal modulation of the sublimation on a nucleus with non-zero obliquity.  The hemisphere facing the Sun on the way to perihelion is not necessarily the hemisphere facing it after perihelion, breaking symmetry between pre- and post-perihelion mass loss rates.  Order of magnitude global asymmetries in CO and CO$_2$ production rates with respect to H$_2$O have been measured in the high obliquity (52\degr) nucleus of 67P/Churyumov-Gerasimenko (\cite{Kel17}, \cite{Hoa19}), where they are apparently correlated, in part, with seasonal illumination differences. Without knowing the pole direction of O3 and the surface distribution of volatiles on the nucleus, we cannot calculate the magnitude of such an effect.  However, we note that the true anomaly changed by a full $\sim$230\degr~between the last pre-perihelion and the first post-perihelion observations plotted in Figure \ref{Cometas_plot}.  On a nucleus with obliquity $\sim$90\degr~and a pole pointed near the Sun at large pre-perihelion $r_H$, the illumination could switch completely from one hemisphere to the other around perihelion.  Seasonal effects can also contribute to the small heliocentric index by causing the sublimation rate to slow down as the comet approaches the Sun as a result of changing illumination.  While qualitatively plausible, we lack knowledge of the nucleus shape, surface inhomogeneity and rotation vector needed to quantitatively account for the fading lightcurve.

Lastly, we note that sublimated material leaving the nucleus exerts a small torque that can change the spin and, if the nucleus is sufficiently small, can drive the nucleus to rotational instability \citep{Jew21}. That this apparently did \textit{not} happen in the case of O3 allows us to set a model-dependent lower limit to the nucleus size.  To do this we note that  rotational spin-up due to outgassing torques can be avoided if  loss of volatiles is confined to a surface layer of thickness 

\begin{equation}
\Delta Z \lesssim \frac{2 r_n^2 \omega}{15 k_T V_{th}}
\label{Deltaz}
\end{equation}

\noindent where $r_n$ is the nucleus radius, $\omega$ its rotational angular frequency, $k_T$ is the dimensionless moment arm and $V_{th}$ is the speed of the outflowing gas \citep{Jew25}.  Rearranging Equation \ref{Deltaz} gives a limit to the radius 

\begin{equation}
r_n \gtrsim \left(\frac{15 k_T V_{th} \Delta Z}{2 \omega} \right)^{1/2}.
\end{equation}

\noindent if a strengthless nucleus is to survive against rotational disruption.  We assume $k_T$ = 0.007 \citep{Jew21}, $V_{th} = 500$ m s$^{-1}$, and adopt $\omega = 1.2\times10^{-4}$ s$^{-1}$, corresponding to the 15 hour median rotational period measured for cometary nuclei \citep{Kok17}.  Substituting $\Delta Z \sim$ 4 m, as above, we find $r_n \gtrsim$ 1.0 km for the nucleus to survive.  This limit is clearly approximate, since $\omega$ and $k_T$ are unmeasured in O3 and, in principle, each could be different from the  assumed values, above,  by factors of several.  However, this lower limit is consistent with the finding that the $\sim$25\% of  LPC nuclei that do not survive perihelion are sub-kilometer bodies \citep{Jew22}.  It is also consistent with the post-perihelion estimate of the nucleus radius, $r_n \lesssim$1.7 km, based on photometry (Section \ref{section3.1}), and points to a nucleus that survived perihelion simply by virtue of being too large to torque to breakup.   

All we know about the Oort cloud is based on inferences from observations of  LPCs entering the planetary region of the solar system.  These comets are subject to rapid physical evolution, rendering the observable population different from the intrinsic population in the cloud.   
   For example, size-dependent destruction processes (rotational disruption is a good example \citep{Jew22}) will flatten the nucleus size distribution of observable  LPCs relative to the ``primordial'' value in the source reservoir. This will affect estimates of both the Oort cloud population and the total mass, potentially by large factors.   The destructive effects of solar heating also alter the brightness and visibility of comets, and hence are implicated in setting the apparent ratio of dynamically new to returning Oort cloud comets  (\cite{Oor51}, \cite{Lev02}). The present observations of O3, which was reported in the literature to have disintegrated but which clearly survived, remind us of the continuing need for a careful observational assessment of cometary evolution both before and after perihelion.

\clearpage

\section{SUMMARY}

We present ground-based and space-based coronagraphic observations of long-period comet C/2021 O3 (Panstarrs), designed to examine its fate at its 0.287 au perihelion.

\begin{itemize}

\item Comet C/2021 O3 survived the 0.287 au perihelion, albeit with a factor of $\sim$25 loss of scattering cross-section.  The post-perihelion cross-section of the comet was $\sim$9 km$^2$, setting a photometric upper limit to the nucleus radius $r_n <$ 1.7 km (albedo 0.04 assumed).  Survival against rotational disruption sets a lower radius limit $r_n \gtrsim$ 1.0 km.

\item The heliocentric variation of the cross-section, $C \propto r_H^{-0.59\pm0.21}$, is too shallow to be consistent with water driven activity across the 2 au to 4 au distance range, but closer to the index expected of a super volatile ice, presumably CO or CO$_2$.

\item Coronagraphic observations in a forward-scattering geometry (phase angles 168\degr~to 177\degr) at perihelion  show a linear morphology with total magnitude $V = 5.4\pm$1.0.  Forward scattering at these large angles is dominated by particles $\sim$9 microns in radius.  

\item There is a real pre- vs.~post-perihelion asymmetry in the activity of O3, but we find no compelling evidence to support published claims that its nucleus was destroyed.  Simple seasonal effects on a high obliquity nucleus offer a plausible explanation for the observed asymmetry.

\end{itemize}

\acknowledgments
We thank the organizers of Cometas and their contributing observers for providing useful photometry on-line.  We thank Jane Luu and the anomymous referee for comments on the manuscript.





\begin{deluxetable}{lcccrrrrrcrrrrl}
\tabletypesize{\scriptsize}
\tablecaption{Observations 
\label{geometry}}
\tablewidth{0pt}
\tablehead{\colhead{UT Date} & \colhead{DOY$_{21}$\tablenotemark{a}} & \colhead{Obs\tablenotemark{b}}   & \colhead{$r_H$\tablenotemark{c}} & \colhead{$\Delta$\tablenotemark{d}}  & \colhead{$\alpha$\tablenotemark{e}} & \colhead{$\theta_{- \odot}$\tablenotemark{f}} & \colhead{$\theta_{-V}$\tablenotemark{g}} & \colhead{$\delta_{\oplus}$\tablenotemark{h}} & \colhead{Mag\tablenotemark{i}}    }

\startdata


2021	Aug	28	&	240	&	B74	&	3.936	&	3.094	&	9.2	&	193.5	&	266.8	&	-8.7	&	18.74	$\pm$	0.02	  \\	
2021	Sep	3	&	246	&	CFHT	&	3.855	&	2.980	&	8.5	&	182.4	&	267.4	&	-7.5	&		18.0 $\pm$	0.1 \\	
2021	Sep	5	&	248	&	B96	&	3.841	&	2.962	&	8.4	&	178.2	&	267.8	&	-7.1	&	18.03	$\pm$	0.61	 \\	
2021	Sep	8	&	251	&	C23	&	3.805	&	2.916	&	8.2	&	171.5	&	268.6	&	-6.4	&	18.68	$\pm$	0.15	 \\	
2021	Sep	15	&	258	&	C23	&	3.721	&	2.822	&	8.0	&	154.3	&	272	&	-4.7	&	18.15	$\pm$	0.11	   \\	
2021	Sep	24	&	267	&	C23	&	3.611	&	2.722	&	8.5	&	132.0	&	281.3	&	-2.2	&	18.03	$\pm$	0.07	 \\	
2021	Oct	4	&	277	&	B74	&	3.487	&	2.641	&	10.1	&	111.5	&	302.7	&	0.8	&	18.03	$\pm$	0.01	   \\	
2021	Oct	7	&	280	&	C23	&	3.449	&	2.623	&	10.7	&	106.5	&	311.5	&	1.8	&	17.84	$\pm$	0.07	  \\	
2021	Oct	15	&	288	&	C23	&	3.348	&	2.587	&	12.6	&	95.6	&	335.1	&	4.3	&	17.64	$\pm$	0.28	 \\	
2021	Oct	24	&	297	&	232	&	3.233	&	2.566	&	14.7	&	86.5	&	353.4	&	7.1	&	18.03	$\pm$	0.00	 \\	
2021	Oct	25	&	298	&	213	&	3.220	&	2.565	&	14.9	&	85.7	&	354.8	&	7.4	&	17.91	$\pm$	0.00	 \\	
2021	Oct	27	&	300	&	C23	&	3.194	&	2.563	&	15.4	&	84.1	&	357.3	&	8	&	17.90	$\pm$	0.07	 \\	

2021	Oct	31	&	304	&	C23	&	3.142	&	2.562	&	16.3	&	81.2	&	1.2	&	9.2	&	17.86	$\pm$	0.04	 \\	
2021	Nov	5	&	309	&	C23	&	3.076	&	2.564	&	17.4	&	78.0	&	4.5	&	10.5	&	17.87	$\pm$	0.05	 \\	
2021	Nov	19	&	323	&	213	&	2.889	&	2.587	&	19.8	&	71.2	&	8	&	13.9	&	17.85	$\pm$	0.00	 \\	
2021	Nov	22	&	326	&	B96	&	2.848	&	2.593	&	20.2	&	70.1	&	8.1	&	14.4	&	17.93	$\pm$	0.08	 \\	
2021	Nov	30	&	334	&	B74	&	2.738	&	2.612	&	21.1	&	67.4	&	7.6	&	15.8	&	17.45	$\pm$	0.03	\\	
2021	Dec	8	&	342	&	213	&	2.625	&	2.629	&	21.6	&	65.1	&	6.2	&	16.9	&	17.95	$\pm$	0.00	 \\	
2021	Dec	12	&	346	&	B74	&	2.568	&	2.636	&	21.8	&	64.1	&	5.3	&	17.3	&	17.70	$\pm$	0.06	 \\	
2021	Dec	14	&	348	&	232	&	2.540	&	2.639	&	21.8	&	63.6	&	4.8	&	17.5	&	17.43	$\pm$	0.18	 \\	
2022	Jan	9	&	374	&	C23	&	2.153	&	2.633	&	20.8	&	58.5	&	355.2	&	18	&	17.12	$\pm$	0.09	 \\	
2022	Jan	24	&	389	&	C23	&	1.915	&	2.574	&	19.0	&	56.6	&	347.7	&	16.9	&	16.62	$\pm$	0.10	 \\	
2022	Apr	27	&	482	&	ST-A	&	0.343	&	0.631	&	163.5	&	148.1	&	120.7	&	3.6	&	  	N/A	  	\\	
2022	Apr	28	&	483	&	ST-A	&	0.361	&	0.606	&	171.6	&	194.2	&	121.9	&	2.3	&	  	N/A	  	\\	
2022	Apr	29	&	484	&	ST-A	&	0.380	&	0.585	&	177.4	&	291.3	&	122.7	&	0.9	&	  	N/A	  	\\	
2022	Apr	30	&	485	&	ST-A	&	0.400	&	0.567	&	172.2	&	303.5	&	123.2	&	-0.6	&	  	N/A	  	\\	
2022	May	1	&	486	&	ST-A	&	0.421	&	0.552	&	165.2	&	307.4	&	123.5	&	-2.1	&	  	N/A	  	\\	
2022	May	10	&	495	&	CA80	&	0.641	&	0.605	&	108.2	&	23.6	&	163.4	&	-66.4	&	15.90	$\pm$	0.50	\\	
2022	May	13	&	498	&	C23	&	0.689	&	0.615	&	101.4	&	21.9	&	164.5	&	-70.6	&	14.55	$\pm$	0.03	 \\	
2022	May	17	&	502	&	JO35	&	0.797	&	0.659	&	87.4	&	24.4	&	168.6	&	-71.7	&	16.14	$\pm$	0.04	\\	
2022	May	20	&	505	&	C23	&	0.843	&	0.685	&	82.2	&	31.0	&	175.6	&	-69.9	&	16.86	$\pm$	0.16	 \\

\enddata


\tablenotetext{a}{Day of Year, 1 = UT 2021 January 1}
\tablenotetext{b}{Observer and telescope details: 213: Ramon Montcabrer, 0.30 m, 232: Esteban Masquefa, 0.25 m, B74: Josep de Montmagastrell, 0.40 m, B96: Erik Bryssinck, 0.4 m, C23: Alfons Olmen, 0.2 m, ST-A = STEREO-A, CA80 = Calar Alto 0.8 m, CFHT = Canada France Hawaii 3.6 m, JO35 =  Jaeger Observatory 0.35 m}
\tablenotetext{c}{Heliocentric distance, in au }
\tablenotetext{d}{Geocentric distance, in au }
\tablenotetext{e}{Phase angle, in degrees }
\tablenotetext{f}{Position angle of projected anti-solar direction, in degrees }
\tablenotetext{g}{Position angle of negative heliocentric velocity vector, in degrees}
\tablenotetext{h}{Angle of observatory from orbital plane, in degrees}
\tablenotetext{i}{Magnitude within a 10\arcsec~radius aperture}

\end{deluxetable}



\clearpage

\clearpage 

\clearpage 

\clearpage

\clearpage 

\clearpage 

\clearpage 

\clearpage 

\clearpage 

%
%

%
%

\end{document}